\pdfoutput=1

\documentclass[11pt]{article}

\usepackage[]{acl}

\usepackage{times}
\usepackage{latexsym}
\usepackage{enumitem}
\usepackage{MnSymbol,wasysym}
\usepackage[T1]{fontenc}

\usepackage[utf8]{inputenc}

\usepackage{microtype}

\usepackage{inconsolata}

\usepackage{graphicx}

\title{OpenOmni: A Collaborative Open Source Tool for \\Building Future-Ready Multimodal Conversational Agents}

\author{
  \textbf{Qiang Sun\textsuperscript{1}},
  \textbf{Yuanyi Luo\textsuperscript{2}},
  \textbf{Sirui Li\textsuperscript{3}},
  \textbf{Wenxiao Zhang\textsuperscript{1}},
  \textbf{Wei Liu\textsuperscript{1}}
\\
  \textsuperscript{1}The University of Western Australia, Perth, WA, Australia,
\\
  \textsuperscript{2}Harbin Institute of Technology, Harbin, China,
  \textsuperscript{3}Murdoch University, Perth, WA, Australia,
\\
  \small{
    \textbf{Correspondence:} \href{mailto:pascal.sun@research.uwa.edu.au}{pascal.sun@research.uwa.edu.au}
  }
}

\begin{document}

\maketitle
\begin{abstract}
Multimodal conversational agents are highly desirable because they offer natural and human-like interaction.
However, there is a lack of comprehensive end-to-end solutions to support collaborative development and benchmarking.
While proprietary systems like GPT-4o and Gemini have demonstrated impressive integration of audio, video, and text with response times of 200-250ms, challenges remain in balancing latency, accuracy, cost, and data privacy.
To better understand and quantify these issues, we developed \textbf{OpenOmni}, an open-source, end-to-end pipeline benchmarking tool that integrates advanced technologies such as Speech-to-Text, Emotion Detection, Retrieval Augmented Generation, Large Language Models, along with the ability to integrate customized models.
OpenOmni supports local and cloud deployment, ensuring data privacy and supporting latency and accuracy benchmarking. 
This flexible framework allows researchers to customize the pipeline, focusing on real bottlenecks and facilitating rapid proof-of-concept development. OpenOmni can significantly enhance applications like indoor assistance for visually impaired individuals, advancing human-computer interaction.
Our demonstration video is available \url{https://www.youtube.com/watch?v=zaSiT3clWqY}, demo is available via \url{https://openomni.ai4wa.com}, code is available via \url{https://github.com/AI4WA/OpenOmniFramework}.

\end{abstract}
\begin{figure*}[ht]
    \centering
    \includegraphics[width=0.78\linewidth]{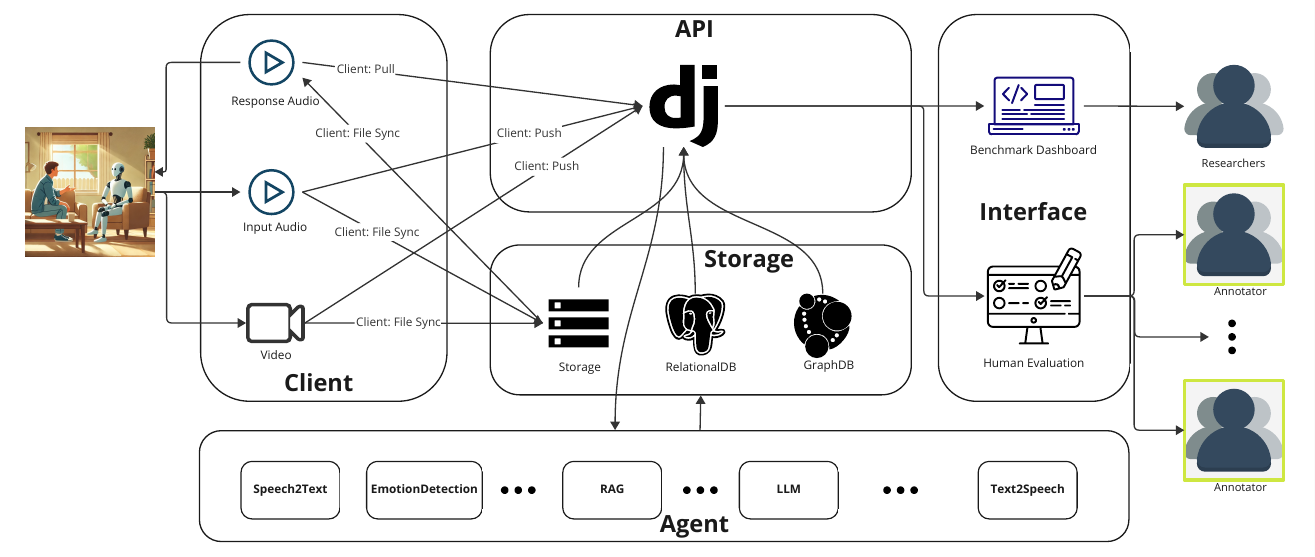}
    \caption{Architecture Design for OpenOmni Framework}
    \label{fig:architecture-design}
\end{figure*}

\section{Introduction}
Large Language Models (LLMs)~\cite{survey-llm2023, survey-llm2024} demonstrated remarkable capabilities in understanding user intentions and following instructions. 
However, text-only human-computer interaction (HCI) is often insufficient~\cite{DBLP:conf/emnlp/ZhangLB23}.
OpenAI recently released their new flagship model, GPT-4o, which can reason across audio, video, and text in real time. 
The impressive performance is achieved with response times between 200-250ms, which is acceptable for large-scale applications\footnote{\url{https://openai.com/index/hello-gpt-4o/}}. 
Google soon followed with their latest multimodal competitors, indicating a clear trend towards multimodal generative models and applications\footnote{\url{https://blog.google/products/gemini/}}. 
LLaVA~\cite{liu2023visualinstructiontuning} is one of the early publicly available solutions for multimodal large models integrating text and images.
However, there is currently no open source end-to-end conversational agents implementation and demonstration publicly available online.

The ideal form of multimodal HCI should mirror human interactions, incorporating video and audio inputs with audio outputs. 
Despite the availability of various modular components, there is no comprehensive integrated open-source implementation to foster research and innovation in this field. 
Integrating existing models, such as audio speech recognition (\textit{Speech2Text}), multimodal large models~(MLMs), and text-to-speech synthesis~(\textit{TTS})—into a multimodal conversation system reveals significant challenges in balancing latency and accuracy. 
Historically, accuracy has been a major hurdle; however, advancements in large language models~(LLMs) have substantially improved contextual relevance. 
The main challenge is reducing end-to-end latency while maintaining accuracy. While OpenAI and Google have shown it’s possible, the open-source community lacks alternatives that replicate this performance.

Another issue is data privacy. 
The GPT-4 family of solutions also raise concerns about cost and data privacy. 
Since GPT-4 is closed-source, users must upload their data to the server via a paid API, raising privacy issues. The privacy policy of GPT\footnote{\url{https://www.gpt.com.au/privacy-policy}} informs users that various forms of personal information, including account details, user content, communication information, and social media data, are collected when users create accounts to access ChatGPT services~\cite{wu2024unveiling}. 

To support the rapid and responsible development of this new HCI format, establishing robust evaluation and benchmarking protocols is essential. For instance, if a user initiates a conversation in a sad and urgent tone, the system should respond appropriately with patience. 
Evaluating this interaction is crucial and challenging for widespread adoption.
Our project aims to bridge these gaps by:
\begin{itemize}[itemsep=0.2pt, parsep=0pt]
    \item Developing an open-source framework for end-to-end customizable conversational agents.
    \item Providing options for a fully local or controllable end-to-end multimodal conversation solution, addressing privacy concerns.
    \item Setting up tools to annotate and benchmark latency and accuracy performance, allowing rapid proof of concept development and research.
\end{itemize}

\noindent To achieve this goal, we propose the \textbf{OpenOmni} framework, an open-source, end-to-end multimodal pipeline that integrates advanced technologies such as Speech-to-Text (Speech2Text), Emotion Detection, Retrieval Augmented Generation~(RAG), Large Language Models~(LLMs), and Text-to-Speech~(TTS). The framework gathers video and audio data from cameras and microphones, processes it through a customizable agents pipeline, and responds via a speaker, as illustrated in Figure~\ref{fig:architecture-design}. 
\textbf{OpenOmni} can be deployed on a local server, ensuring secure data management and addressing privacy concerns. 

For research purposes, it includes tools for easy annotation and benchmarking, offering real-time monitoring and performance evaluation of latency. 
Users can annotate individual components and entire conversations, generating comprehensive benchmark reports to identify bottlenecks. 
The open-source nature of OpenOmni allows for adaptation across different application domains, such as aged care, personal assistant, etc. 
Each pipeline component can be enabled or disabled based on specific use cases, facilitating flexible and efficient deployment. 
Additionally, the framework supports the easy addition of extra models, enabling comparisons and further experimentation.
The \textbf{OpenOmni} framework allows researchers to focus on solving critical bottlenecks without reinventing the wheel, fostering innovation in multimodal conversational agents. 
It enables rapid proof-of-concept development, such as indoor conversational robots assisting visually impaired individuals.

\section{Related works}
\noindent\textbf{Solution options}
\begin{figure}[ht]
    \centering
    \includegraphics[width=0.9\linewidth]{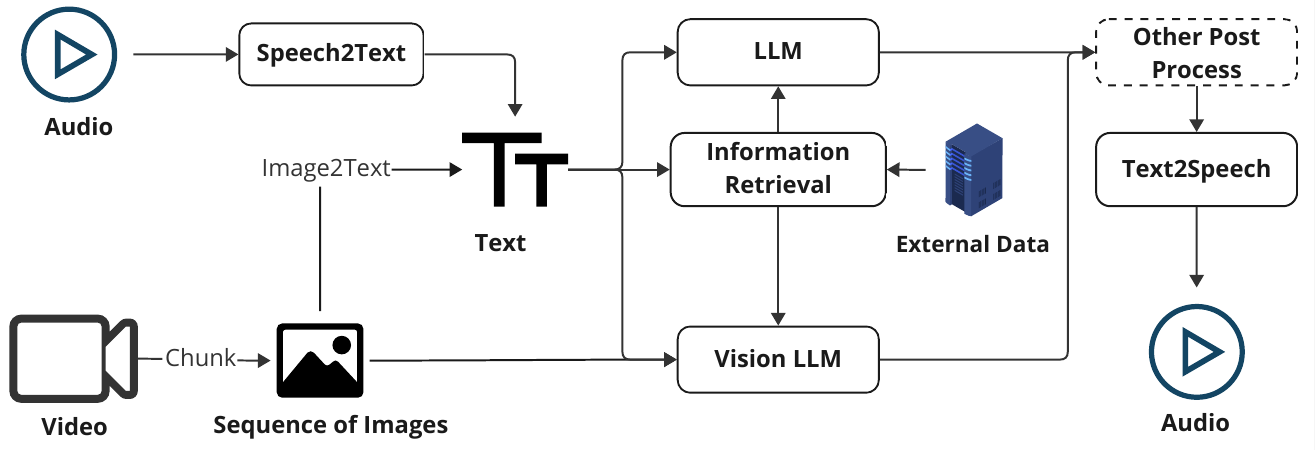}
    \caption{Traditional \textbf{\textit{divide-and-conquer}} end-to-end multimodal conversation system}
    \label{fig:traditional_module}
\end{figure}
Traditional end-to-end multimodal conversation systems, as shown in Figure~\ref{fig:traditional_module}, typically use a \textbf{\textit{divide-and-conquer}} strategy, splitting the process into sub-tasks: speech-to-text (automatic speech recognition), image-to-text, text generation, and text-to-speech~\cite{9864589}. Speech-to-text converts spoken language into text, while image-to-text generates textual descriptions of images. 
Text generation, often powered by large language models, produces contextually appropriate responses, and text-to-speech converts these responses back into spoken language. 
These core components form the backbone of the conversational pipeline. 
Image-to-text adds essential context, enhancing natural human-computer interaction, and additional tasks like emotion detection tailor responses to the user's emotional state. 
A safe guard module can optionally be integrated to ensure responses are appropriate, non-harmful, and controllable, maintaining interaction integrity, especially in sensitive scenarios.
While this modular approach allows for optimization of individual components, the accumulated latency and accuracy errors can render the end-to-end system impractical for real-world use.

\begin{figure}[ht]
    \centering
    \includegraphics[width=0.9\linewidth]{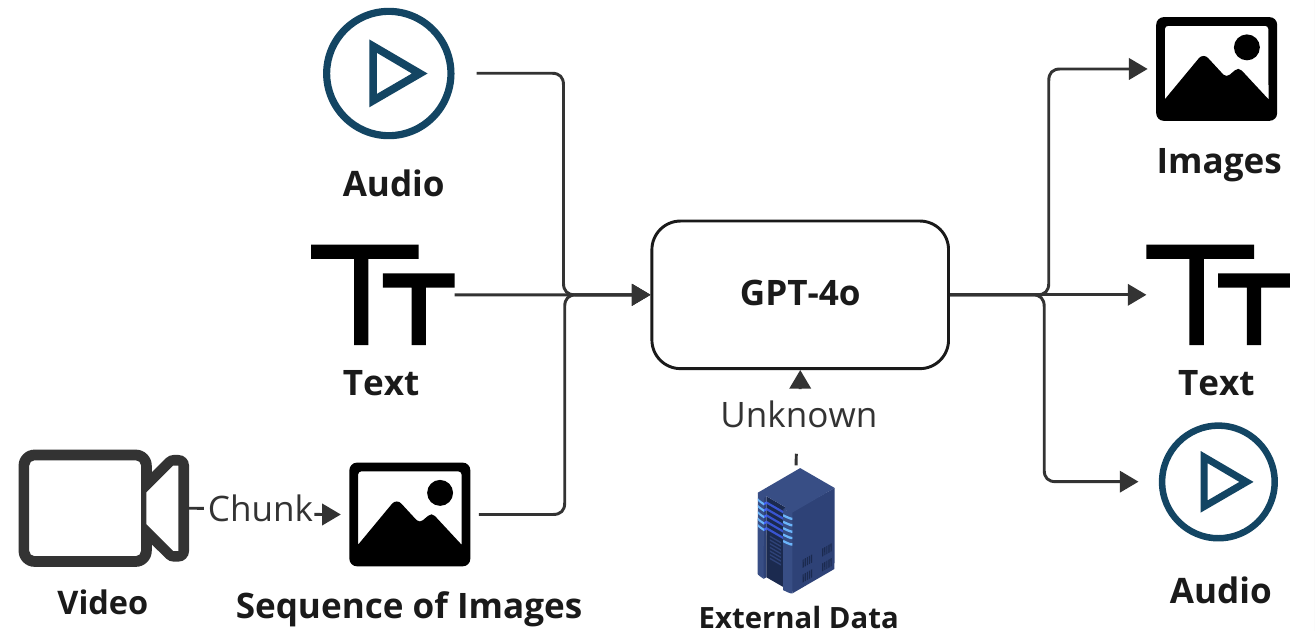}
    \caption{Our assumptions about how the \textbf{\textit{fully end-to-end}} model: GPT-4o works}
    \label{fig:gpt4o_workflow}
\end{figure}

While GPT-4o is marketed as a \textbf{\textit{fully end-to-end}} model, where inputs are video, audio or texts and outputs are audio, images or text, its technical details are unreleased. 
We assume, as shown in Figure~\ref{fig:gpt4o_workflow}, that audio and video frames are fed into modules generating text, audio, and image outputs.
The demonstration video suggests GPT-4o has memory capabilities, but specifics and limitations are unclear. 
It is also unknown if the system can directly integrate external private data.

Unlike the \textbf{\textit{divide-and-conquer}} approach, a \textbf{\textit{fully end-to-end}} neural network can incorporate more contextual information, such as tone, multiple speakers, and background noises, resulting in more flexible outputs. 
This approach can theoretically reduce latency by eliminating orchestration bottlenecks.
However, both solutions face significant challenges due to immense data I/O, especially from video. Video files are large, straining servers and models, increasing computational costs, and causing latency from data transfer and model inference. 
Real-time conversation requires streaming processing, posing further latency challenges. 
In OpenAI's demonstration\footnote{\url{https://www.youtube.com/watch?v=RI-BxtCx32s}}, a USB-C connection to an iPhone was used to ensure a \textbf{stable} internet connection, highlighting these issues.

\begin{figure}[ht]
    \centering
    \includegraphics[width=0.9\linewidth]{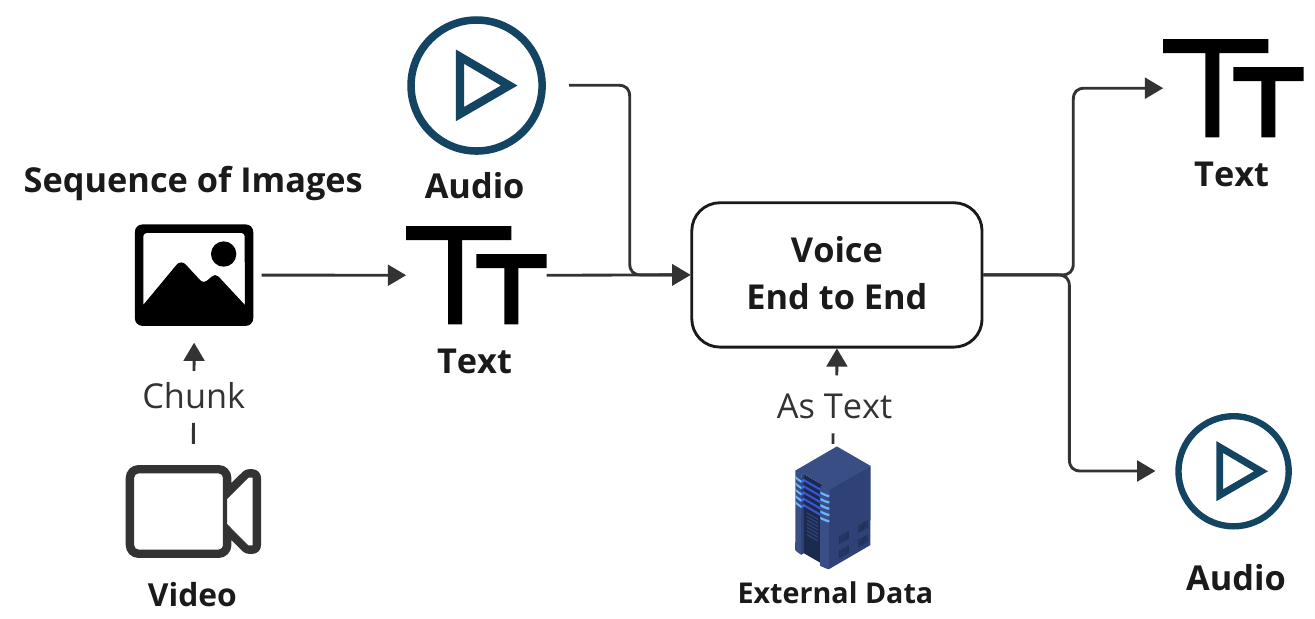}
    \caption{\textbf{\textit{Hybrid}} solution via the combination of \textit{image2text} and end-to-end voice model}
    \label{fig:hybrid_solution}
\end{figure}

Recently, Kyutai, a technology company from France, released a planned open-source, fully end-to-end multimodal conversational AI called \textit{Moshi}~\footnote{\url{https://kyutai.org/}}. 
This model supports text and audio modalities, excluding images, and claims to achieve an end-to-end latency of 200ms. 
We can integrate the video modality via an \textit{Image2Text}~\cite{Lin_2021_CVPR} module into \textit{Moshi}, creating a \textbf{\textit{Hybrid}} solution, as shown in Figure~\ref{fig:hybrid_solution}, that combines the \textit{divide-and-conquer} and \textit{fully end-to-end} approaches.
Another feasible \textbf{\textit{Hybrid}} solution is to use speech-to-text to convert audio into text, then feed this text along with video (processed into image sequences) to a vision language model, which generates text responses. 
These responses can then be processed through text-to-speech, as illustrated in Figure~\ref{fig:traditional_module} via the \textit{Vision LLM} line.
\begin{figure}[ht]
    \centering
    \includegraphics[width=0.6\linewidth]{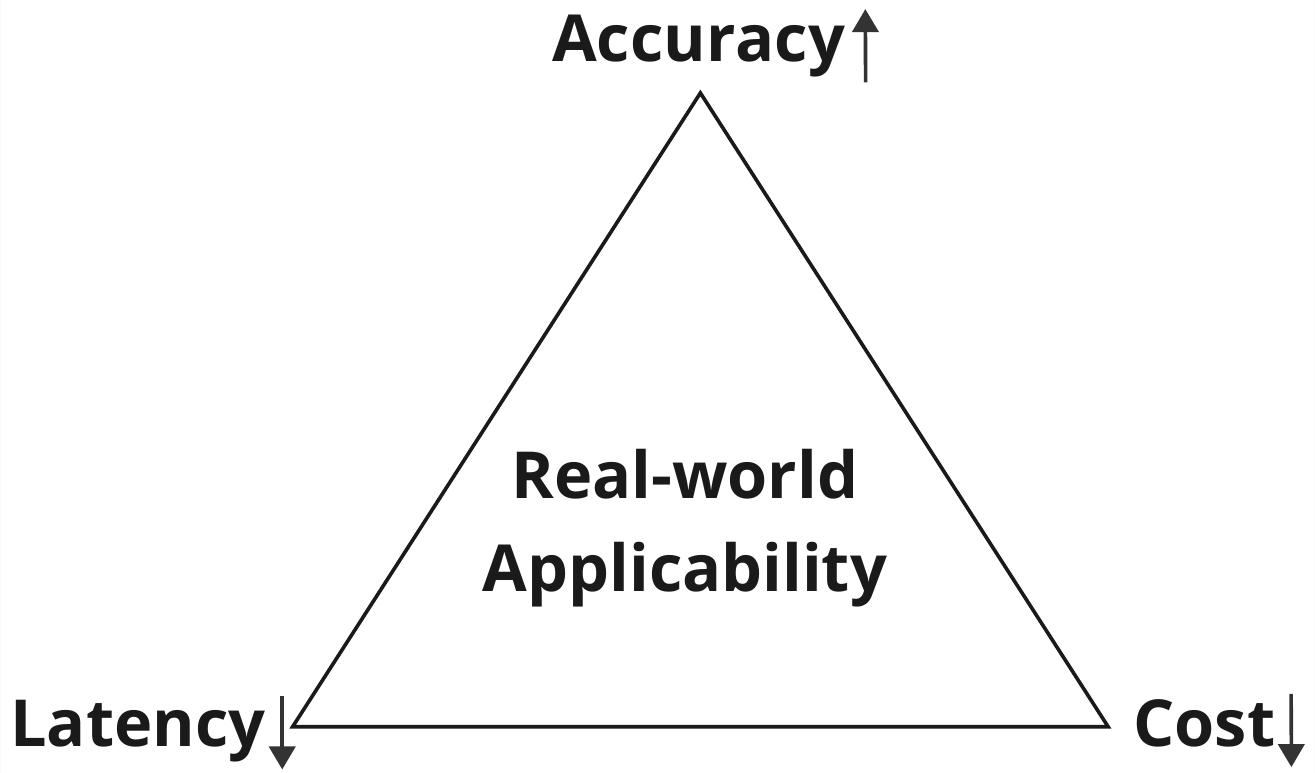}
    \caption{Constraint triangle for real-world applicability in multimodal conversational agent development}
    \label{fig:Triangle}
\end{figure}
Multimodal end-to-end conversational agents, like OpenAI’s GPT-4, show promise, but large-scale application is challenging due to the need to balance latency, accuracy, and cost. 
Generating real-time responses between 200-400 ms is difficult. As shown in Figure~\ref{fig:Triangle}, the primary goal is to reduce latency and cost while improving accuracy, enhancing the real-world applicability of conversational agents.

\noindent\textbf{Evaluation metrics}

To ensure efficient and effective collaboration, consistent and comparable evaluation metrics are essential. For speech-to-text, the Word Error Rate (WER)~\cite{roy2021semanticwerunifiedmetricevaluation} measures transcription accuracy, with a lower WER indicating better performance. Text-to-speech evaluation includes objective metrics like the Mean Opinion Score (MOS)~\cite{Streijl2016MeanOS} for naturalness and intelligibility, and the Signal-to-Noise Ratio (SNR)~\cite{1709898} for clarity, as well as subjective human ratings. Text generation is the most challenging to evaluate, using metrics like BLEU, ROUGE, and METEOR~\cite{EVTIKHIEV2023111741}, which compare generated text to references but may not fully capture response quality and relevance. Evaluating text generation often requires large-scale datasets, which are not always available.
These metrics are widely adopted by the research community, including OpenAI. However, real-world applications require evaluation in production environments, considering diverse factors beyond these metrics. For instance, an aged care conversational agent should avoid sensitive topics that may be specific to each individual. Subjective opinions vary by region, highlighting the need for customizable and innovative automatic or semi-automatic evaluation approaches for conversational agents.

\section{System design}


\subsection{Requirement analysis}

The system receives audio and video input, produces audio as the output. 
Initially, we need two modules: one to collect audio and video data from the microphone and camera, and another to play audio through a speaker. 
These \textbf{Client} modules should support diverse devices, such as a smartphone, a laptop, or a Raspberry Pi. 
The collected data will then be fed to a server.

The server, referred to as \textbf{API}, should manage audio, video data, and metadata. It should have access to a storage layer that includes a relational database, file management, and a graph database for potential GraphRAG integration. While the \textbf{API} can reside on the same instance as the \textbf{Client} module, we prefer them to be separate for better adaptability. This separation introduces the challenge of \textbf{sharing large volumes of data between modules}.
If the \textbf{API} is cloud-based, the audio and video data need to be uploaded to the cloud, for example using AWS S3, Azure Blob Storage, or Google Cloud Storage.
However, the upload process can become a bottleneck, making the data transfer time-consuming. If the server is local, within the same network as the \textbf{Client}, transfer latency will be reduced. However, this setup requires running the large language model locally, addressing data ownership and privacy concerns but potentially increasing model inference latency and compromising accuracy due to limited computing resources.
Another solution is edge computing, where video data is pre-processed on edge devices and summarized for the \textbf{API}. While this can be a research direction, data compression may cause information loss and reduce end-to-end performance.

The pipeline components will need modification if developers want to adopt the framework and integrate with their work. To ensure flexibility, this part should be an independent module that can run locally or in the cloud. Researchers and developers should be able to easily integrate new components into this \textbf{Agent} module, further challenging the sharing of large datasets between modules.

Lastly, we want to generate benchmarks to understand the latency and accuracy performance of the entire pipeline. For tasks that are hard to evaluate automatically, such as determining the appropriateness of the LLM response, we propose and develop an annotation module to allow human annotators to easily evaluate results and generate benchmark reports.

\subsection{System architecture}
Based on the requirements, we designed our system as shown in Figure~\ref{fig:architecture-design}. The system is divided into five modules: \textbf{Client}, \textbf{API}, \textbf{Storage}, \textbf{User Interface}, and \textbf{Agent}, all primarily developed in Python.
The \textbf{Client} module includes two submodules: the \textbf{Listener} for collecting video and audio data, and the \textbf{Responder} for playing audio. The \textbf{Storage} module consists of file storage for media, a relational database (PostgreSQL) for metadata, and a graph database (Neo4j) for potential GraphRAG integration.
The \textbf{API} module, built with the Django framework, extends Django's admin interface and permission control system to develop the benchmark and annotation interface. Django's maturity and large support community make it ideal for production development.
The \textbf{Agent} module, also in Python, includes all agent related submodules, allowing deployment on suitable compute nodes without altering the architecture. Communication between the \textbf{Client}, \textbf{API}, and \textbf{Agent} modules will be via RESTful endpoints.
For \textbf{sharing large data between modules}, local deployments (e.g., \textbf{Client} on Raspberry Pi, \textbf{API} and \textbf{Agent} on local servers) will use FTP for file synchronization. In cloud solutions (e.g., AWS), files will be uploaded to AWS S3\footnote{\url{https://aws.amazon.com/s3/}}, triggering a Lambda function to download files to an AWS Elastic File Storage~(EFS)~\footnote{\url{https://aws.amazon.com/efs/}} shared by the \textbf{API} and \textbf{Agent} modules.
We use \textit{Docker} and \textit{Docker Compose} to manage all modules, allowing easy setup with a single {\tt docker compose up} command.

\section{Demonstration}

\subsection{Datasets}
Most multimodal question answering datasets focus on multiple-choice questions rather than open-ended conversations~\cite{sundar-heck-2022-multimodal}. 
Some, like Image-Chat~\cite{DBLP:journals/corr/abs-1811-00945}, involve multimodal conversations with images as extra input, but the output is often multiple-choice or text-based~\cite{survey-multi-modal-dialogue}.
A major hurdle in developing multimodal conversational agents is the lack of appropriate datasets.

While there is no shortage of data from human-human interactions or extracted from movies and YouTube videos, we lack efficient methods to organize this data into structured datasets.
For specific domain applications, collecting data from human interactions and extracting datasets to train systems would be beneficial, allowing the agents to mimic human behavior.
Our OpenOmni Framework provides both capabilities: extracting conversational datasets from videos and testing them through the pipeline to evaluate agents' responses, or collecting data from real-world scenarios to generate datasets for further research.

\subsection{Can ``AI'' be your president?}
One intensive conversational scenario is a debate. We extracted segments from the US Presidential Debate 2024 between Biden and Trump\footnote{\url{https://www.youtube.com/watch?v=-v-8wJkmwBY}}, focusing on Biden addressing the public and handling questions. 
After downloading the videos, you can use a prepared script in our codebase to split them into segments. This script allows you to specify the start and end times of each conversation, enabling you to create a conversational dataset from the videos.
These segments were fed into our pipeline to evaluate its performance under different configurations: OpenAI Whisper for speech-to-text, GPT-4o vision model, and text-to-speech (GPT4O\_ETE); a locally deployed quantization LLM with Whisper, text-to-speech, and our emotion detection model for video input~(QuantizationLLM\_ETE); a version using HuggingFace LLM for inference (HF\_ETE); and a version using only Whisper, GPT-3.5, and text-to-speech, ignoring the video modality (GPT35\_ETE).
We ran the \textbf{Agent} modules on an NVIDIA-3080 GPU with 12GB memory.

\begin{figure}[ht]
    \centering
    \includegraphics[width=0.99\linewidth]{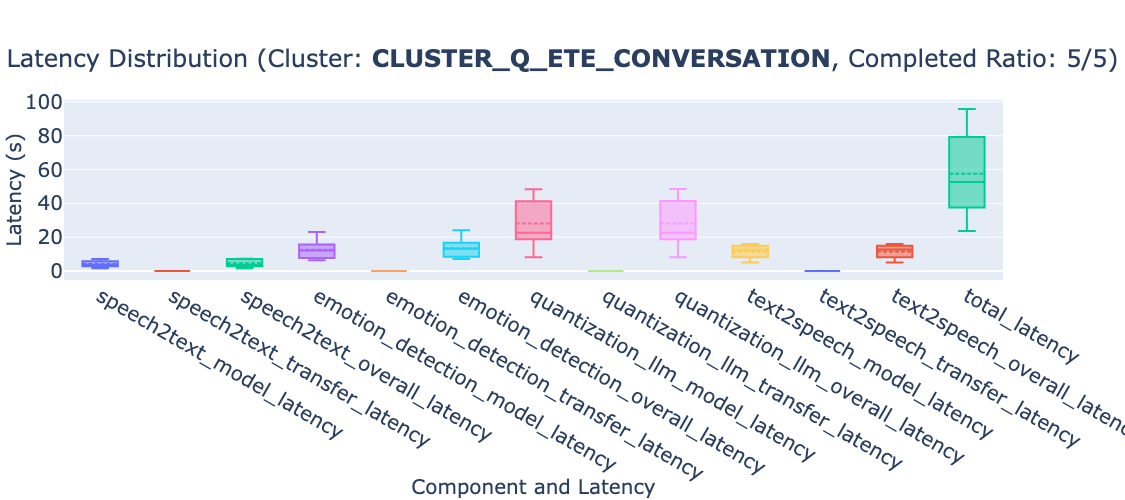}
    \caption{Screenshot of the end-to-end latency benchmark statistics for the setup: \textit{Local Whisper, Emotion Detection, Quantization LLM, and OpenAI Text-to-Speech}. This visualization is one example of the generated benchmark report; you can customize it or explore more details within our demo.}
    \label{fig:quantization_llm}
\end{figure}

The latency benchmark statistics are automatically generated. For example, the GPT4O\_ETE configuration has an average latency of 45 seconds, with the GPT-4o vision model accounting for 31 seconds.
The fastest configuration is GPT35\_ETE, averaging around 15 seconds, with most of the time consumed by the text-to-speech part, because the generated content is quite long and comprehensive.
The slowest configuration is HF\_ETE, taking around 189 seconds, with the LLM model inference step taking the longest time.
QuantizationLLM\_ETE takes an average of 60 seconds, as shown in Figure~\ref{fig:quantization_llm}, with the LLM model inference averaging 28 seconds and our emotion detection model averaging around 10 seconds.

\begin{figure}[ht]
    \centering
    \includegraphics[width=0.99\linewidth]{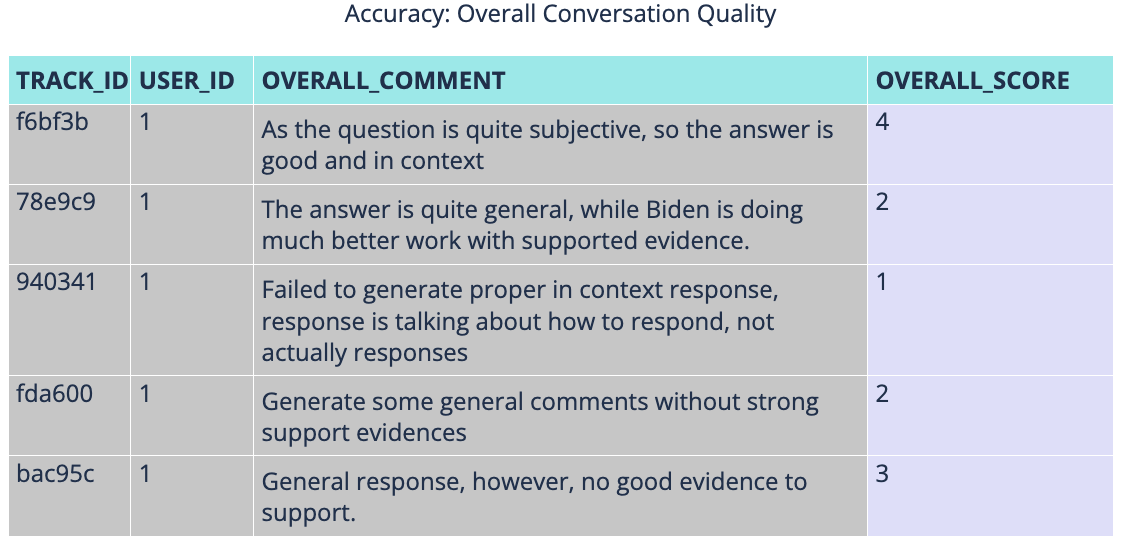}
    \caption{Screenshot of annotated overall conversation accuracy statistics and comments for each conversation within GPT4O\_ETE. Scores range from 0 to 5.}
    \label{fig:gpt4oaccuracy}
\end{figure}
After annotation with our interface, accuracy statistics are automatically generated. 
The accuracy metrics here include evaluation metrics like WER, CER~\cite{roy2021semanticwerunifiedmetricevaluation} for speech2text task, overall scores given by the annotators, etc.
As shown in Figure~\ref{fig:gpt4oaccuracy}, the average score for each conversation is 2.4. Text-to-speech can be improved with more natural emotion or personality. The generated content is often too general and sometimes inappropriate. Biden's responses are more in-context and evidence-supported. The pipeline excelled only in answering a subjective question about Biden's age, where the GPT-4o pipeline performed well.
The GPT35\_ETE pipeline had the best overall accuracy, but its responses were often in-context yet pompous. Thus, Biden still outperforms AI. In conclusion, ``AI cannot be the President of the US just yet, considering both latency and accuracy.''\smiley{}

\subsection{Assist the visually impaired}
\begin{figure}[ht]
    \centering
    \includegraphics[width=0.99\linewidth]{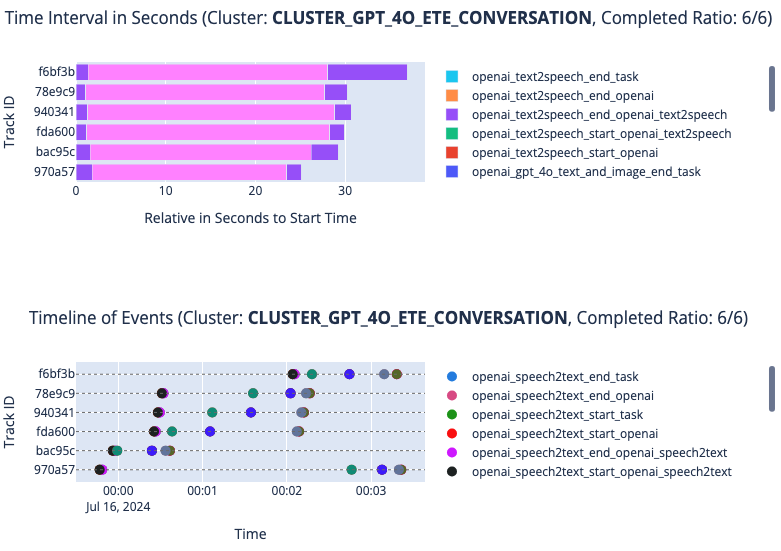}
    \caption{Screenshot visualizing detailed latency benchmark information for each conversation round}
    \label{fig:gpt-4o-assist-latency}
\end{figure}

While latency and the need for external information currently preventing AI from mission critical tasks, conversational agents can be production-ready and useful for non-latency-critical areas that do not require extensive external knowledge. Assisting indoor activities for the visually impaired is one such application, in which you can either utilize high-speed internet or limit data transfer to local exchanges. 
These type applications can benefit from maintaining high input/output rates, helping to mitigate latency issues.
We prepared questions for the visually impaired, including locating objects, navigating indoors, and inquiries about the surroundings. Six questions were sampled and fed to the GPT4O\_ETE pipeline. 
One scenario demonstration is included in our provided YouTube video.
In this scenario, video and audio data stream from the client side and are saved to storage along with exportable metadata accessible via the admin portal. This setup allows you to export annotated datasets, including raw video and audio data, for developing new models.
The latency statistics in Figure~\ref{fig:gpt-4o-assist-latency} show responses within approximately 30 seconds.

Annotated results show a 4.7/5 accuracy, but the agent lacks specific skills for assisting the visually impaired. For example, ideally, it should provide step-by-step instructions on grabbing a coffee cup rather than just a general description.
This indicates that while conversational agents are nearly ready for assisting the visually impaired with indoor activities, improvements in latency and response quality are still needed.

\section{Conclusion}

Multimodal conversational agents offers a more natural human-computer interaction, exemplified by models like GPT-4o. However, real-world constraints necessitate balancing cost, latency, and accuracy, which may explain why GPT-4o's full capabilities are not yet accessible. 

There are several technical options to achieve this, including traditional \textbf{\textit{divide-and-conquer}} methods, \textbf{\textit{fully end-to-end}} models like GPT-4o, and \textbf{\textit{Hybrid}} approaches. The \textbf{\textit{fully end-to-end}} approach inherently allows for lower latency, while the \textbf{\textit{divide-and-conquer}} method faces latency issues when coordinating multiple components. Both approaches must address the challenge of handling large data I/O. If models are deployed locally, local network I/O issues can be more manageable. However, OpenAI's models are closed-source, making local deployment impractical. While deploying other vision models locally is feasible, achieving high accuracy may be limited by local computational resources. \textbf{\textit{Hybrid}} solutions provides alternative approaches: pre-processing or compressing large data locally and then utilizing cloud-based models, or converting video to text and integrating it into the end-to-end voice model.

We developed the OpenOmni framework to enable researchers to integrate their work into an end-to-end pipeline. The framework supports various solutions, allows for pipeline customization, generates latency performance reports, and provides an annotation interface for accuracy review. These features facilitate the creation of benchmark reports to identify and address key issues.

Testing with the US Presidential debate scenario highlighted latency as a critical issue, particularly with large video data. Integrating external knowledge remains a challenge, emphasizing the need for efficient Retrieval-Augmented Generation~(RAG). For applications like indoor assistance for the visually impaired, latency improvements and model adaptation are both essential.

The OpenOmni framework can significantly benefit the research community by facilitating the collection and management of new datasets, integrating various conversational agents approaches, and generating automatic latency benchmarks. Its annotation interface aids in accuracy performance review, making OpenOmni production-ready for suitable application scenarios and fostering further development in multimodal conversational agents.

\bibliography{custom}

\end{document}